\documentclass[journal]{IEEEtran}
\usepackage{amsmath,amsfonts}
\usepackage{algorithmic}
\usepackage{algorithm}
\usepackage{array}
\usepackage[caption=false,font=normalsize,labelfont=sf,textfont=sf]{subfig}
\usepackage{textcomp}
\usepackage{stfloats}
\usepackage{url}
\usepackage{verbatim}
\usepackage{graphicx}
\usepackage{cite}
\usepackage{multirow}
\usepackage{booktabs}
\usepackage{float}
\usepackage{makecell}
\usepackage{amssymb}

\newtheorem{lemma}{Lemma}
\begin{document}
	
	\title{Reliability Assessment of Power System Based on the Dichotomy Method}
	
	\author{
		\thanks{This work was supported by the National Key Research and Development Program of China (No. 2023YFA1011302)}
		\IEEEauthorblockN{ Wenjie Wan, Han Hu, Feiyu Chen\IEEEauthorrefmark{1}, Xiaoyu Liu, Kequan Zhao}
		\thanks{Corresponding Author: Feiyu Chen \quad Email: fchen\_cqnu@163.com}
		\thanks{W. Wan, Han Hu, F. Chen and X. Liu are with the National Center for Applied Mathematics in Chongqing,  Chongqing Normal University, China.}
		\thanks{K. Zhao is with the School of Mathematical Sciences, Chongqing Normal University, China.}
	}
	
	\maketitle
	
	\begin{abstract}
		With a sustainable increase in the scale of power system, the number of states in the state space grows exponentially, and the reliability assessment of the power system faces enormous challenges. 
		Traditional state-by-state assessment methods, such as state enumeration (SE) and Monte Carlo simulation (MCS) methods, have encountered performance bottlenecks in terms of efficiency and accuracy. 
		In this paper, the Boolean lattice representation theory of the state space was studied, and a dichotomy method was proposed to efficiently partition the state space into some disjoint sub-lattices with a relatively small number of optimal power flow (OPF) operations. Based on lattice partition, the reliability indices of the entire space can be calculated lattice-by-lattice. In addition, alone with the partitioning procedure, the calculated loss of load probability (LOLP) monotonically increases and rapidly tends to the analytic value with the designated error bound. Moreover, we designed a customized Monte Carlo sampling method in lattices of interest to compute expected energy not supply (EENS). 
		The experiments are conducted on the RBTS and RTS-79 systems. The results show that the proposed method achieves the analytic LOLP of the RBTS system after five hundreds of OPF operations, which is about hundreds of times faster than traditional methods, and the designed Monte Carlo sampling method converged after thousands of OPF operations on test systems.
	\end{abstract}
	
	
	\section*{Notation}
	\addcontentsline{toc}{chapter}{Notation}
	
	\begin{tabular}{ll}
		\(S\) & State space \\	
		\(N\)  & Normal region \\
		\(F\) & Failed region \\
		\(\mathcal{L}\) & Lattice \\
		\(\mathcal{F}\) & The set of failed Lattices \\
		\(\mathcal{M}\) & The set of mixed Lattices \\
		\(s\) & System state \\
		\(x_i\)  & Status of component \(i\) \\
		\(a_i\)  & Availability of component \(i\) \\
		\(q_i\) & Unavailability of component \(i\) \\
		\(P\) & Probability function \\
		\(I\) & Impact function \\
		\(\Phi\) & Structure function \\
		\(C\) & Optimal load shedding function \\
		\(\hat{0}\) & Minimum element \\
		\(\hat{1}\) & Maximum element \\
		\(\left[ s_1 , s_2 \right]\) & Interval \\
	\end{tabular}
	
	\section{Introduction}
	Reliability assessment is a critical part of the stable operation of power systems \cite{Billinton1994}. In recent years, the rapid increase in new energy devices, such as wind and photovoltaic systems, has led to a significant expansion in the scale of power systems. This expansion poses a considerable challenge to the reliability of the operation of the power system\cite{10711297}.
	
	Reliability indices are used to assess the ability of a power system to meet the continuous power supply demands of consumers under specified conditions \cite{Billinton1994}. The reliability assessment methods can be broadly categorized into three types: the state enumeration method (SE), the Monte Carlo simulation method (MCS) and the hybrid method. 
	
	The SE method, which is an analytical method, calculates reliability indices by individually evaluating each state in the state space of the system. However, the number of states increases exponentially with respect to the number of components. Therefore, SE becomes computationally infeasible to obtain accurate reliability indices for large-scale composite systems. Various methods such as state space truncation \cite{sharaf1988reliability}, contingency selection algorithm \cite{mikolinnas1981advanced}, fast contingency screening technique \cite{LIU20081019,6523187}, analytical model and solution method \cite{8362973}, state expansion \cite{852155} and minimum cut sets \cite{5342441} have been developed to improve computing efficiency. These methods reduce the number of Optimal Power Flow (OPF) calculations to some extent. Nevertheless, they often overlook a significant number of high-level states. To address this limitation, the impact increment method \cite{7741639} converts the traditional SE formula to an impact increment version. The Lagrange multiplier method was employed to establish the relationship between optimal load shedding and state \cite{9294070}. This approach allows directly calculating the optimal load shedding for a large number of states, thus significantly reducing the computational burden.
	
	MCS approximates the reliability indices by estimating the expected value of a number of randomly sampled states, where the number of samples usually does not depend on the size or complexity of the system \cite{8094959}. However, due to the unique characteristics of the state space of the system, which encompasses a large number of high-probability states with no load shedding, the efficiency of MCS is significantly dependent on the stability of the system\cite{7581049}. In highly stable systems, the convergence rate of MCS is extremely low. Techniques such as regression variable \cite{4310363}, stratified sampling \cite{151823}, state space pruning \cite{575787}, importance sampling \cite{141713}, the cross-entropy method \cite{8684335} and the MPLP-TLSD method \cite{8510885} have been proposed to accelerate convergence rate.
	
	The hybrid methods, which first partition the entire state space into two regions and then perform SE and MCS methods on respective region, integrate the strengths of both SE and MCS. Several methods have been developed, including hybrid methods \cite{Clancy1983ProbabiliticFF,141702}, hybrid methods based on space decomposition \cite{doi:10.1049/iet-gtd.2009.0281}, and hybrid methods based on impact increments \cite{9215203}. Although these methods have proven effective, they inevitably sample states with no load shedding. Considering the probability of a state, load shedding is likely to occur in states with low probabilities \cite{doi:10.1049/iet-gtd.2009.0281}.
	
	The existing SE methods and variants are difficult to enumerate the large number of high-level contingency states, but those low-probability contingency states contribute to calculating the reliability indices. In addition, the MCS method and its derivative methods suffer from slow convergence. The need for a large number of samples to achieve acceptable accuracy may result in considerable computational time. Moreover, the inherent random sampling mechanism leads to the uncertainty of the evaluated result.
	
	Both the SE and MCS methods, which evaluate system reliability state-by-state, have encountered performance bottlenecks in terms of efficiency and accuracy. In order to improve the efficiency and accuracy of the assessment, we propose a novel reliability assessment method based on lattice partition of state space. The entire state space was partitioned into disjoint sub-lattices, and the reliability indices of the system can be calculated lattice-by-lattice. More specifically, based on the lattice representation and Cartesian product of state space, a dichotomy method was proposed to divide a lattice into two sub-lattices with an equal number of states. By skillfully selecting components and lattices, the dichotomy method efficiently partitions the entire space into union of disjoint sub-lattices with different characteristics after a relatively small number of OPF operations.  Alone with the dichotomy procedure, the  calculated LOLP, which is the total probabilities of failed lattices, monotonically increases and rapidly tends to the analytic value with the designated error bound. Moreover, the EENS is calculated using MCS in the failed region, which is the union of failed sub-lattices (FMCS). Unlike the direct MCS method, the proposed FMCS circumvents the sampling of low-level states with no load shedding and achieves higher convergence rates, and the convergence rate is independent of the scale and stability of the power system. The effectiveness of the proposed approach is demonstrated through numerical experiments on the RBTS and RTS-79 systems, showcasing significant improvements in computational efficiency and accuracy compared to SE and MCS. In conclusion, the main contributions of this paper are as follows.
	\begin{enumerate}
		\item A novel lattice-by-lattice reliability assessment framework based on state space partition was proposed.   
		\item An efficient dichotomy method was proposed to partition state space into disjoint sub-lattices after a relatively small number of OPF operations.
		\item An analytic formulation was given to calculate LOLP progressively and monotonically; a customized MCS method was proposed to calculate EENS with high convergence rate.        
	\end{enumerate}
	
	\section{Physical model and mathematical model}
	Electrical components, such as generators, transformers, and transmission lines, are fundamental units of a power system. A component is considered operational when it performs its designated function within specified technical parameters. In contrast, a failed component denotes the termination of its ability to perform its designated function. The status of a component is commonly represented as a random variable \(x_i\).
	
	\begin{equation}
		x_i = \begin{cases} 
			0, & \text{if the component is operational} \\
			1, & \text{if the component is failed}
		\end{cases}
	\end{equation}
	
	The reliability of a component is characterized by its failure rate function \(\lambda(t)\) and its repair rate function \(\mu(t)\). The lifetime and repair time of a component are assumed to follow exponential distributions, then failure and repair rates become constants, denoted \(\lambda\) and \(\mu\), respectively. Consequently, the availability \(a\) and the unavailability \(q\) of the component are defined as:
	\begin{equation}
		a = \frac{\mu}{\lambda + \mu}, \quad q = \frac{\lambda}{\lambda + \mu}
	\end{equation}
	
	A system consists of multiple components interconnected to serve a specific production purpose. The reliability of the system is determined by both the reliability of its components and its overall topological connection. A typical monotonic system exhibits the following properties:
	\begin{enumerate} 
		\item Each component in the system has two possible statuses: operational or failed.
		\item The system itself, composed of these components, is assumed to have two statuses: normal or failed.
		\item The system is normal if all the components are operational.
		\item The system is failed if all the components are failed.
		\item In a failed system, the failure of additional components does not restore functionality. In contrast, in a normal system, the operation of an additional component does not cause the system to fail. 
	\end{enumerate}
	
	A state consisting of \(n\) components can be represented as:
	\begin{equation}
		s = (x_1, x_2, \dots, x_n)
	\end{equation}
	where \( x_i \in \{0, 1\} \). Each component \( i \) has two statuses: 0 and 1. Thus, the total number of states is \( 2^n \). The set of all possible states is called the state space, denoted as \( S \), where each element in \( S \) is a state represented by an \( n \)-dimensional vector, i.e.,
	\begin{equation}
		S = \left\{ s \in \mathbb{R}^n \mid s(i) = 0 \text{ or } 1 \right\}
	\end{equation}
	
	Assuming the components are independent, the \(i\)-th component in use can have two statuses with availability \(a_i\) and unavailability \(q_i\). The probability of each state \(s\) is expressed as:
	\begin{equation}\label{prop_state}
		P(s) = \left( \prod_{k \in U_s} a_k \right) \cdot \left( \prod_{l \in D_s} q_l\right)
	\end{equation}
	where \(U_s\) is the set of components that are operational in the state \(s\), and \(D_s\) is the set of components that have failed in the state \(s\).
	
	In this paper, a failed state refers to the occurrence of load shedding, while a normal state refers to the absence of load shedding. The load shedding function is determined using the DC optimal power flow model (OPF):
	
	\textbf {Objective function:}
	\begin{equation}
		C(s) = \min \sum_{i \in ND} C_i
	\end{equation}
	\textbf {Constraints:}
	\begin{equation}
		\begin{gathered}
			T(s) = A(s) (P_G - P_D + C) \\
			\sum_{i \in NG} P_{G_i} + \sum_{i \in ND} C_i = \sum_{i \in ND} P_{D_i} \\
			P_G^{\text{min}} \leq P_G \leq P_G^{\text{max}} \\
			0 \leq C_i \leq P_D \\
			|T(s)| \leq T^{\text{max}}
		\end{gathered}
	\end{equation}
	where \( C(s) \) is the load shedding amount, \( T(s) \) is the active power vector of the lines in state \( s \), \( A(s) \) is the incidence matrix that represents the relationship between the active power vector of the lines and the node injection power vector in state \( s \). \( P_G \), \( P_D \), and \( C \) are the node generator active power injection vector, node active load vector, and node load reduction vector, respectively. \( NG \) and \( ND \) are the sets of generator buses and load buses.
	
	Define the structure function \(\Phi: S \rightarrow \{0, 1\}\) on the state space, where the output of the structure function represents the status of state, i.e.,
	\begin{equation}\label{structure function}
		\Phi(s) =
		\begin{cases}
			0,~~ & \text{if} ~C(s)=0\\
			1,~~ & \text{if} ~C(s)>0
		\end{cases}
	\end{equation}

	The reliability index can be expressed as follows:
	
	\begin{equation}
		R = \sum_{s \in S} P(s) I(s)
	\end{equation}
	where \( P(s) \) is the probability of the occurrence of state \( s \), and \( I(s) \) is the impact function of the state \( s \). If \( I(s) = \Phi(s)\), Index \(R\) is LOLP. If \( I(s) = C(s)\), Index \(R\) is EENS.
	
	\section{The Lattice Representation Theory of the State Space}
	\subsection{The State Space and Boolean Lattice}
	The state space of an \( n \)-component system \( S \) is a partially ordered vector set, where the partial order relation "\(\leq\)"  is defined as follows: for any \(s_1, s_2 \in S\), \(s_1 \leq s_2\) if and only if \(s_1(i) \leq s_2(i)\), for \(i = 1, 2, \dots, n\). Based on the definition of lattice, it follows that \( S \) constitutes a \( n \)-dimensional Boolean lattice $\mathcal{L}$ \cite{Bruce}, the minimum element of the lattice \( \hat{0} = (0, 0, \dots, 0) \) represents the state in which all components are operational (a 0-level state), and the maximum element of the lattice \( \hat{1} = (1, 1, \dots, 1) \) represents the state in which all components are failed (a \( n \)-level state). In other words, we have \( \hat{0} \leq s \leq \hat{1} \) for any \( s \in \mathcal{L} \). Thus, the lattice $\mathcal{L}$ can be represented as a closed interval:
	\begin{equation}
		\mathcal{L} = [\hat{0}, \hat{1}]
	\end{equation}
	
	For two states \( s_1, s_2 \in S \) with \( s_1 \leq s_2 \), the closed interval \([s_1, s_2]\) also forms a Boolean lattice with a minimum element $s_1$ and a maximum element $s_2$. Specifically, if \( s_1 = s_2 \), the closed interval \([s_1, s_1]\) constitutes a 0-dimensional Boolean lattice with only one element.
	
	In addition, let \(\mathcal{L}^{\ast}=[s_1,s_2] \subseteq [\hat{0}, \hat{1}]\) be a sub-lattice of $\mathcal{L}$, then the corresponding vector set $S^{\ast}$ of sub-lattice $\mathcal{L}^{\ast}$ also forms a subsystem, which consists of the components $i$ with $s_1(i) = 0, s_2(i) = 1$. The components with $s_1(i) = s_2(i) = 1$ must fail and the components with $s_1(i) = s_2(i) = 0$ must be operational in the subsystem. 
	
	Therefore, we claim that the state space of a \( n \)-component system is equivalent to a \( n \)-dimensional lattice.
	
	\subsection{Lattice Partition of the State Space}
	
	The properties of the monotonic system imply that the structure function (\ref{structure function}) of the state space is an order-preserving function\cite{Bruce}. According to the structure function, the entire state space \( S \) can be divided into two disjoint regions. The normal region, denoted as \( N \), is the set of all normal states; and the failed region, denoted as \( F \), is the set of all failed states. 
	\begin{equation}
		N = \{ s \in S \mid \Phi(s) = 0 \},\ F = \{ s \in S \mid \Phi(s) = 1 \}.
	\end{equation}
	
	The lattice can be categorized into three different types: normal lattice, failed lattice, and mixed lattice, where the states of the normal lattice are normal, the states of the failed lattice are failed, and the mixed lattice contains normal and failed states. In addition, the characteristic of a lattice can be determined by the status of its minimum and maximum elements.
	\begin{lemma}\label{characteristic of a lattice}
		Let \(\mathcal{L} = [s_1,s_2]\) be a lattice, and  $\Phi$ is the structure function, then 
		\begin{itemize}
			\item $\mathcal{L}$ is a normal lattice if \( \Phi(s_2) = 0 \).
			\item $\mathcal{L}$ is a failed lattice if \( \Phi(s_1) = 1 \).
			\item $\mathcal{L}$ is a mixed lattice if \( \Phi(s_1) = 0 \) and \( \Phi(s_1) = 1 \).
		\end{itemize}		
	\end{lemma}
	
	Note that each state can be considered as a sub-lattice of 0-dimension, the normal region \( N \) can be partitioned into the union of disjoint normal lattices \(\{\mathcal{L}_i^N\}\), and the failed region \(F\) can be partitioned into the union of disjoint failed lattices \(\{\mathcal{L}_j^F\}\). Thus, the state space \( S \) can be partitioned as:
	
	\begin{equation} \label{lattice partition}
		S = N \cup F = \left( \bigcup_i \mathcal{L}_i^N \right) \cup \left( \bigcup_j \mathcal{L}_j^F \right)
	\end{equation}
	
	Clearly, the lattice partition of state space is not unique.
	
	\subsection{Reliability Index reformulation based on Lattice Partition}
	Let \( \mathcal{L} = \left[ \hat{0}_\mathcal{L}, \hat{1}_\mathcal{L} \right] \) be a lattice, the probability and impact of \( \mathcal{L} \) can be defined as follows.
	
	The probability of the lattice is defined as the sum of the probabilities of all states in the lattice, expressed as:
	\begin{equation}\label{prop_lattice}
		\begin{aligned}
			P(\mathcal{L}) = \sum_{s \in \mathcal{L} }P(s) = \left( \prod_{k \in U_\mathcal{L}} a_k \right) \cdot \left( \prod_{l \in D_\mathcal{L}} q_l \right)
		\end{aligned}
	\end{equation}
	where \( U_\mathcal{L} \) is the set of components that are operational in state \( \hat{1}_\mathcal{L} \), and \( D_\mathcal{L} \) is the set of components that are failed in state \( \hat{0}_\mathcal{L} \). 	If \( \hat{0}_\mathcal{L} = \hat{1}_\mathcal{L} \), the probability of the 0-dimensional lattice (\ref{prop_lattice}) coincides with the probability of a single state (\ref{prop_state}).
	
	The impact of the lattice \( I(\mathcal{L}) \) is defined as the expected impact of all states in the lattice, expressed as:
	\begin{equation}
		\begin{aligned}
			I(\mathcal{L}) &= \mathbf{E}_p(I(s)) = \frac{1}{P(\mathcal{L})} \sum_{s \in \mathcal{L}} P(s) I(s)
		\end{aligned}
	\end{equation}
	where $P(\mathcal{L})$ is the probability of $\mathcal{L}$.
	
	Given \(\{\mathcal{L}_i^N\}\) and \( \{ \mathcal{L}_j^F \} \) as a partition of \( S \) in (\ref{lattice partition}), based on the definition of probability and impact of lattice, the reliability index can be rewritten as:
	\begin{equation}\label{reliability index}
		\begin{aligned}
			R = \sum_{s \in S} P(s) I(s) = \sum_j P(\mathcal{L}_j^F) I(\mathcal{L}_j^F)
		\end{aligned}
	\end{equation}
	
	Clearly, the impact of the normal lattice $I(\mathcal{L}_i^N)$ is zero, since the impact of each individual normal state is always zero. This implies that only failed lattices contribute to the reliability index. 
	
	If \( I(s) = \Phi(s)\), the analytical value of LOLP is:
	\begin{equation}
		\begin{aligned}
			\text{LOLP}_{\ast} = \sum_{j} P(\mathcal{L}_j^F) = 1 - \sum_{i} P(\mathcal{L}_i^N)
		\end{aligned}
	\end{equation}
	If \( I(s) = C(s)\), the analytical value of EENS is: 
	\begin{equation}
		\begin{aligned}
			\text{EENS}_{\ast} = \sum_{j} P(\mathcal{L}_j^F) C(\mathcal{L}_j^F)
		\end{aligned}
	\end{equation}

	Once the state space is partitioned as the union of disjoint failed and normal lattices, the reliability index can be calculated in a lattice-by-lattice manner.
	
	\section{The Dichotomy Method for State Space partition}
	In this section, an efficient dichotomy method is proposed to partition the state space into failed and normal lattices.
	\subsection{The Dichotomy Method}
	
	Let $\mathcal{L}$ be a n-dimensional lattice of a n-component system \( S \), the Cartesian product expression of $\mathcal{L}$ is given as:
	\begin{equation}
		\begin{aligned}
			\mathcal{L} & = \left\{ (x_1, x_2, \dots, x_n) \mid x_i \in \{0, 1\}, \, i = 1, 2, \dots, n \right\}\\
			& = \{0, 1\} \times \{0, 1\} \times \cdots \times \{0, 1\} 
		\end{aligned}
	\end{equation}
	where the Cartesian product operation satisfies the distributive law over the union operation:
	\begin{equation}
		\begin{aligned}
			A \times (B \cup C) = (A \times B) \cup (A \times C) \\
			\quad (B \cup C) \times A = (B \times A) \cup (C \times A)
		\end{aligned}
	\end{equation}
	
	For example, the state space a 3-component system can be represented as:
	\begin{equation}
		\begin{aligned}
			\mathcal{L} &  = \left\{
			\begin{aligned}
				& (0, 0, 0), (1, 0, 0), (0, 1, 0), (0, 0, 1), \\
				& (1, 1, 0), (1, 0, 1), (0, 1, 1), (1, 1, 1)
			\end{aligned}
			\right\}\\
			& = \{0, 1\} \times \{0, 1\} \times \{0, 1\}
		\end{aligned}
	\end{equation}

	The partition with component \( i \) of a lattice \(\mathcal{L}\) can be expressed as:
	\begin{equation}
		\begin{aligned}
			\mathcal{L} & = \{0, 1\} \times \{0, 1\} \times \cdots \times \{0, 1\} \\
			& = \{0, 1\} \times \cdots \{0\} \times \cdots \{0, 1\}  \\
			& \: \cup \{0, 1\} \times \cdots \{1\} \times \cdots \{0, 1\} \\
			& = \underline{\mathcal{L}}_i \cup \overline{\mathcal{L}}_i
		\end{aligned}
	\end{equation}
	where \(\underline{\mathcal{L}}_i\) is the set of states that the \(i\)-th component is operational, and \( \overline{\mathcal{L}}_i \) is the set of states that the \(i\)-th component is failed. Fig.1 illustrates an example of partition a 5-dimensional lattice with one component.    
	
	\begin{figure}[t]
		\centering
		\includegraphics[width=2.5in]{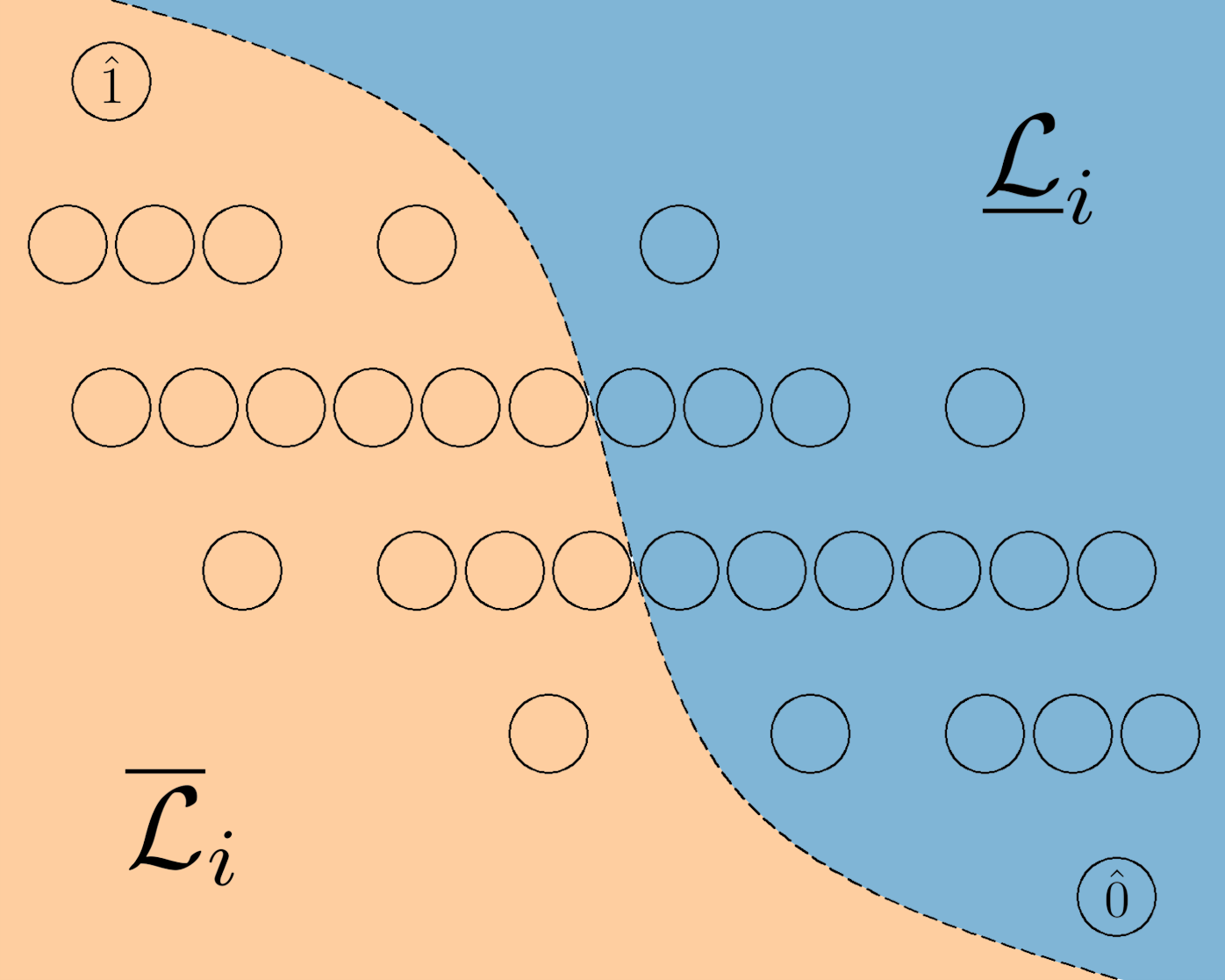}
		\caption{Partition a 5-dimensional lattice with one component.}
		\label{fig_1}
	\end{figure}
	
	According to Lemma \ref{characteristic of a lattice}, the characteristics of sublattices \(\underline{\mathcal{L}}_i\) and \( \overline{\mathcal{L}}_i \) can be determined by the statuses of their minimum and maximum elements. Thus, if the partition procedure is repeated until there are no more mixed lattices, the state space will be partitioned into the union of failed and normal lattices.  Since this partition divides a lattice into two disjoint sublattices with an equal number of states, we refer to this procedure as the dichotomy method for state space partition.
	
	Note that the procedure should be different when partitioning a lattice with different components. As shown in Fig.2, a 5-dimensional lattice is partitioned into the union of failed and normal lattices in two different ways. Although the final normal and failed regions are exactly the same, the number and dimension of sub-lattices varies.    
	
	\begin{figure}[!ht]
		\centering
		\subfloat[]{\includegraphics[width=1.6in]{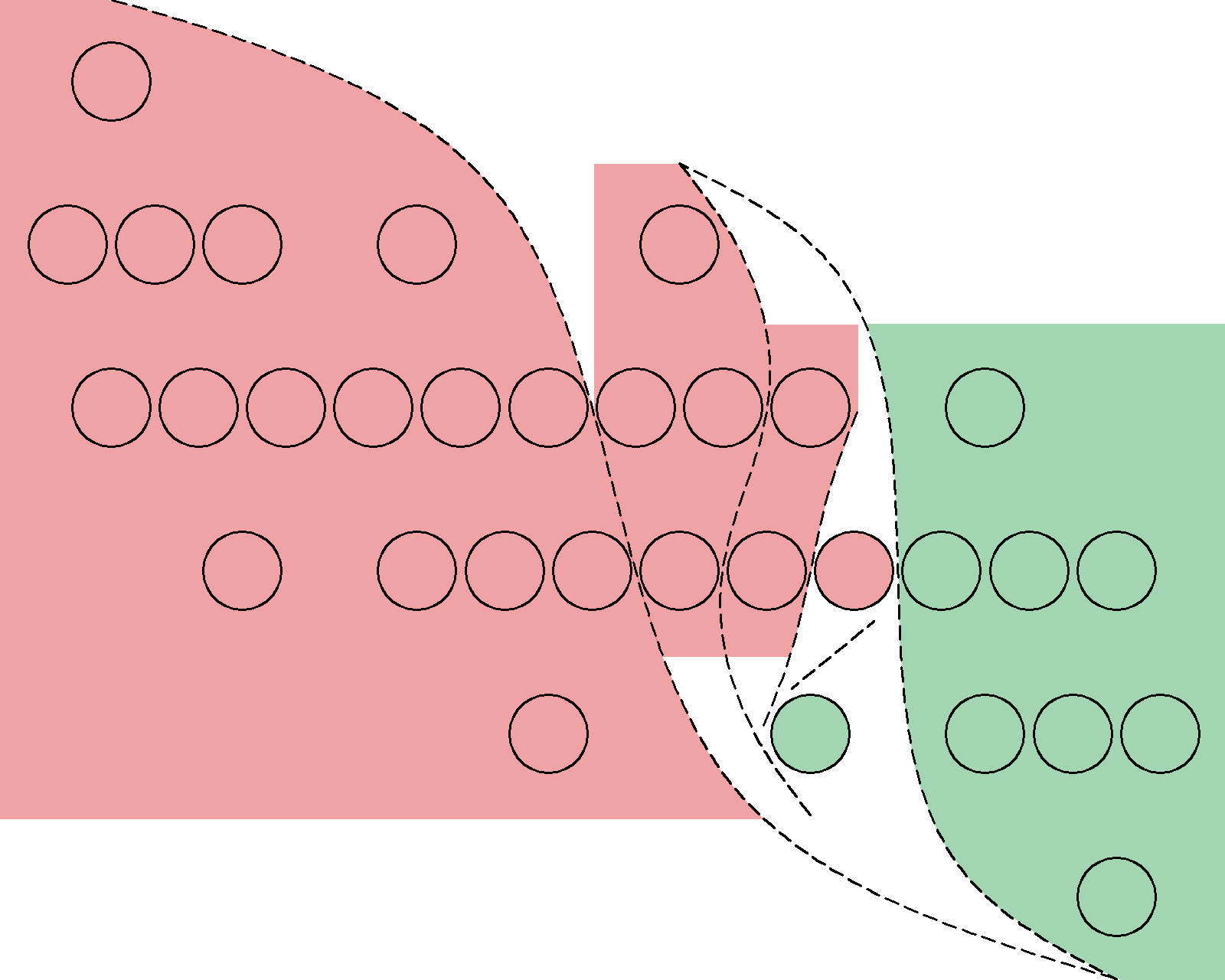}
			\label{partition1}}
		\hfil
		\subfloat[]{\includegraphics[width=1.6in]{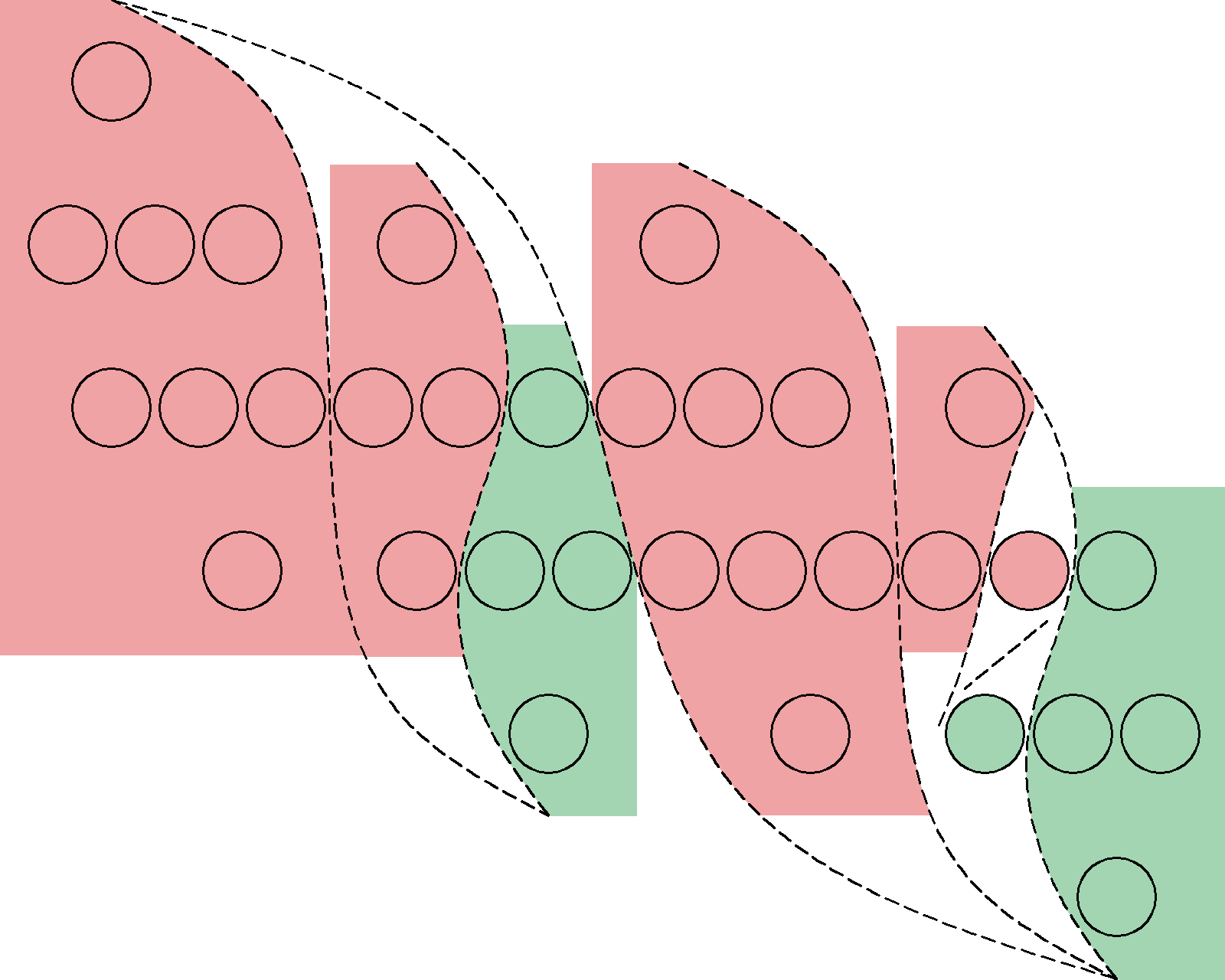}
			\label{partition2}}
		\caption{Different procedures of partitioning the state space into the union of failed (red) and normal (green) lattices.}
		\label{fig2}
	\end{figure}
	
	During the procedure of dichotomy method for space partition, a huge amount of mixed lattices with different components emerged. The automatic selection strategies of the mixed lattice and its component for partitioning are the key points of the efficiency of the dichotomy method. According to (\ref{reliability index}), the reliability indices are only determined by the failed lattices. Therefore, partitioning out failed lattices with a larger probability is beneficial to rapidly approaching the analytical reliability indices.

	\subsection{The selection strategies of lattice and component}
	
	To estimate the failure probability of a lattice, we consider the lattice as a combination of the transmission system and the generation system.
	
	The transmission system can be represented as an undirected graph $\mathcal{G} = \{\mathcal{V},\mathcal{E}\}$, where $\mathcal{V}$ is the set of nodes that represent the buses in the transmission system, and $\mathcal{E}$ is the set of edges that represent the transmission lines between buses. For example, the edge $e= (u,v)$ represents the transmission line between nodes $u$ and $v$, and nodes $u$ and $v$ are adjacent if and only if $(u,v) \in \mathcal{E}$, and node $v$ is a neighbor of node $u$. 
	
	Denote the neighborhood $\mathcal{N}(u)$ as the set of all neighbors of node $u$, and the degree $d(u)$ as the number of neighbors of node $u$. 
	
	The minimum degree node of a graph $\mathcal{G}$ is defined as: 
	\begin{equation}
		v_{\ast} \triangleq  \arg\min_{v \in \mathcal{V}} d(v)
	\end{equation}
	where the minimum degree node $v_{\ast}$ is the bus with the greatest weakness of the transmission system, since there are the fewest lines that connect it. 
	
	The maximum probability that the transmission system fails can be approximated as the probability that all lines connecting to the node $v_{\ast}$ fail:
	
	\begin{equation}
		q_t = \prod_{e \in \mathcal{E}_{\ast}} q(e)
	\end{equation}
	where $\mathcal{E}_\ast$ is the set of all edges that contain node $v_{\ast}$, and $q(e)$ is the unavailability of the line $e$.
	
	Suppose that the generation system consists of $t$ generators $\{g_1,...,g_t\}$ arranged in descending order of installed capacity $c_i$, and the adequacy of the generation system is $a$.
	
	The maximum probability that the generation system fails is the probability that the first $t_{\ast}$ generators fail:
	
	\begin{equation}
		q_g = \prod_{i=1}^{t_{\ast}} q(g_i)
	\end{equation}
	where $q(g_i)$ is the unavailability of the $i$-th generator, and $t_{\ast}$ is defined as:
	\begin{equation}
		t_{\ast} \triangleq \min  \quad \tau \quad \text{s.t.} \quad \sum_{i=1}^{\tau} c_i - a \ge 0
	\end{equation}

        The importance index of a lattice is defined based on its lattice probability and the probability of system failure:
	\begin{equation}\label{importance index}
		\pi( \mathcal{L} ) = P( \mathcal{L} ) \cdot \max (q_t, q_g)
	\end{equation}
	
	After selecting the lattice with the largest importance index, the component for partition is selected based on the value of $q_t$ and $q_g$:
	\begin{itemize}
		\item Select the line $e_\ast = \arg \max_{e \in \mathcal{E}_{\ast}} q(e)$, if $q_t > q_g$.
		\item Select the generator $g_1$, if $q_t \leq q_g$.
	\end{itemize}
	
	\subsection{Stopping criteria and Algorithm of Dichotomy}
    
	For large-scale systems, it is impossible to complete the partition of the entire state space, therefore establishing the stopping criteria is essential for the dichotomy method. 
	\subsubsection{The number of OPF evaluation}
	Considering the limited computing resources, the number of OPF evaluations can be used as a stopping criterion.
	\subsubsection{The total probability of mixed lattices}
	The total probability of the remaining mixed lattices can be used as a stopping criterion. In fact, this probability is the upper bound of the error of the evaluated LOLP index.
	\subsubsection{The average probability of failed lattices}
        During the partition procedure, some failed lattices are divided from the mixed lattices, but their probability continuously decreases due to the use of the importance index for mixed lattice selection. Therefore, the average probability of failed lattices can be used as a stopping criterion. This criterion is related to the growth rate of the evaluated LOLP index.

	The procedure of the dichotomy method for state space partitioning is described in Algorithm 1. It mainly consists of three iterative steps: 1) Select a lattice and a component for partitioning. 2) Evaluate the status of the minimum element of the partitioned lattice. 3) Update the set of failed and mixed lattices.
	\begin{algorithm}[H]
		\caption{The Dichotomy Method for State Space Partition}\label{alg:alg1}
		\begin{algorithmic}
			\STATE \textbf{Input:} $n$-component system \(S\)
			\STATE Initialize the set of failed and mixed lattices: \\
			\( \qquad \qquad \qquad \qquad \mathcal{F} = \emptyset, \quad \mathcal{M} = \{S\} \)
			\REPEAT
			\STATE Select the most important lattice \(\mathcal{L}  \in \mathcal{M}\) by (\ref{importance index}) 
			\STATE Select the most important component \(i\) in \(\mathcal{L}\)
			\STATE Partition \(\mathcal{L}\) with component \(i\) into \(\underline{\mathcal{L}}_i\) and \(\overline{\mathcal{L}}_i\) by (21)
			\STATE Calculate \(C(\hat{0}_{\overline{\mathcal{L}}_i})\) by OPF  
			\STATE \textbf{if} \(C(\hat{0}_{\overline{\mathcal{L}}_i}) > 0\) 
			\STATE \hspace{0.5cm} \(\mathcal{F} = \mathcal{F} \cup \{\overline{\mathcal{L}}_i\}, \mathcal{M} = \mathcal{M} \setminus \{\mathcal{L}\} \cup 
			\{\underline{\mathcal{L}}_i\}\)
			\STATE \textbf{else}
			\STATE \hspace{0.5cm} \(\mathcal{M} = \mathcal{M} \setminus \{\mathcal{L}\} \cup \{\underline{\mathcal{L}}_i\} \cup \{\overline{\mathcal{L}}_i\} \)
			\STATE \textbf{end}
			\UNTIL{ Stopping criteria is met}
			\STATE \textbf{Output:} The set of failed and mixed lattices \(\mathcal{F}\), \(\mathcal{M}\)
		\end{algorithmic}
		\label{alg1}
	\end{algorithm}
	
	\section{Reliability Assessment Based on Lattice Partition}

        After partitioning the state space using the dichotomy method, the failed region $F$ can be represented by the set of failed lattices \(\mathcal{F}\), where the representation error depends on the stopping criteria.
        
        Therefore, the reliability index can be approximated by a lattice-by-lattice manner:
        \begin{equation}\label{reliability index 2}
		R = \sum_{\mathcal{L} \in \mathcal{F}} P(\mathcal{L}) I(\mathcal{L})
	\end{equation}

	\subsection{Calculation of LOLP}
	The LOLP reliability index is the probability of \(\mathcal{F}\).
	\begin{equation}\label{LOLP}
		\text{LOLP} = P(\mathcal{F}) = \sum_{\mathcal{L} \in \mathcal{F}} P(\mathcal{L})
	\end{equation}
	where \(P(\mathcal{L})\) is calculated by (13). 

    With an increasing number of failed lattices, the calculated LOLP monotonically increases with the analytic LOLP as the upper bound. And the gap between the calculated and the analytical LOLP can be controlled by a designed stopping criterion. Therefore, the dichotomy method provides a series of lower bounds of the analytical LOLP during the partitioning procedure. In particular, the dichotomy method achieves the analytical LOLP as the entire space is partitioned.

	\begin{table*}[hb]
		\centering
		\caption{RELIABILITY ASSESSMENT RESULTS OF SE, MCS, and D-FMCS (RBTS)}
		\begin{tabular}{ccccccccc}
			\toprule
			\multicolumn{2}{c}{Method} & \(\beta_{\text{EENS}}\) &  LOLP & LOLP error & EENS & EENS error & Evaluation Number & CPU time(s) \\
			\midrule
			\multicolumn{2}{c}{SE} & -- & 0.94739\% & 0.0000\% & 0.11676 & 0.0000\% & 1,048,576 & 64,100.9 \\
			\midrule
			\multicolumn{2}{c}{\multirow{3}*{MCS}} 
			& 0.1  & 0.84243\% & 11.079\% & 0.09147 & 21.658\% & 20,061 & 1,228.9 \\
			& & 0.05  & 0.92486\% & 2.3774\% & 0.12081 & 3.4659\% & 69,740 & 4,297.6 \\
			& & 0.01  & 0.94738\% & 0.0008\% & 0.11631 & 0.3880\% & 1,781,334 & 109,673.1 \\
			\midrule
			\multirow{4}{*}{D-FMCS}& Dichotomy(\(\text{D}_7\))
			& -- & 0.94738\% & 0.0007\% & -- & -- & 190 & 13.4 \\ 
			\cmidrule{2-9}
			&\multirow{3}*{FMCS}
			& 0.1 & -- & -- & 0.12857 & 10.118\% & 63 & 3.9 \\
			& & 0.05 & -- & -- & 0.11987 & 2.6636\% & 262 & 16.2 \\
			& & 0.01 & -- & -- & 0.11692 & 0.1382\% & 6,874 & 428.2 \\
			\midrule
			\multirow{4}{*}{D-FMCS} & Dichotomy(\(\text{D}_9\))
			& -- & 0.94739\% & 0.0001\% & -- & -- & 578 & 38.3 \\ 
			\cmidrule{2-9}
			&\multirow{3}*{FMCS}
			& 0.1 & -- & -- & 0.12040 & 3.1158\% & 48 & 3.1 \\
			& & 0.05 & -- & -- & 0.11111 & 4.8391\% & 272 & 17.1 \\
			& & 0.01 & -- & -- &0.11765 & 0.7654\% & 6,578 & 408.2 \\
			\bottomrule
		\end{tabular}
	\end{table*}
	\subsection{Estimation of EENS}
        Unlike the calculation of LOLP, it is impossible to calculate the analytic EENS even if the failed region is decomposed as the union of failed lattices, since the impact of each individual failed state is different. MCS is a commonly used method for the estimation of EENS.
        
	The estimated EENS reliability index is given by:
         \begin{equation}\label{EENS}
		\text{EENS} = \text{LOLP} \times \frac{1}{K} \sum_{k=1}^{K} C(s_k)
	\end{equation}
	where $s_k$ are randomly sampled from $\mathcal{F}$, and $K$ is the total number of samples. 

        The above formula provides a good approximation of analytic EENS when $K$ is large enough.
        \begin{equation}
		\begin{aligned}
			\text{EENS} &= \text{LOLP} \times \frac{1}{K} \sum_{k=1}^{K} C(s_k) \\
			&\approx P(\mathcal{F}) \times \sum_{s \in \mathcal{F}} \frac{P(s)}{P(\mathcal{F})}C(s) \\ 
			&\approx \sum_{s \in F} P(s)C(s) = \text{EENS}_\ast
		\end{aligned}  
	\end{equation}	

    Fortunately, the combination expression of $\mathcal{F}$ is beneficial for the design of the sampling strategy. Firstly, a failed lattice \(\mathcal{L}\) is randomly selected from $\mathcal{F}$ based on its probability, then a state $s$ randomly selected from $\mathcal{L}$ based on its probability. The Monte Carlo sampling method designed in the failed region is referred to as FMCS method. 
    
    The procedure of reliability assessment based on lattice partition is described in Algorithm 2, we use the coefficient of variation for EENS as the stopping criterion.    
	
	\begin{algorithm}[H]
		\caption{Reliability Assessment Based on Lattice Partition}\label{alg:alg2}
		\begin{algorithmic}
			\STATE \begin{tabular}{@{}l@{}}
				\textbf{Input:} \(\mathcal{F}\), Convergence threshold \(\beta_{\text{EENS}}\)\end{tabular}
			\STATE Compute LOLP by (\ref{LOLP})
			\REPEAT
			\STATE Randomly select a lattice \(\mathcal{L}  \in \mathcal{F}\) based on its probability
			\STATE Randomly select a state \(s \in \mathcal{L}\) based on its probability
			\STATE Calculate \(C(s)\) by OPF
			\STATE Compute EENS by (\ref{EENS})
			\STATE Calculate \(\beta\) for EENS
			\UNTIL{\(\beta < \beta_{\text{EENS}}\)}
			\STATE \textbf{Output:} LOLP, EENS
		\end{algorithmic}
		\label{alg2}
	\end{algorithm}
	
	\section{Numerical Results}
	In this section, the load models are all based on peak load. The computational platform is Windows 10 64-bit, equipped with a 12th Generation Intel(R) Core(TM) i5-12400F processor running at 2.50 GHz, coupled with 16GB of RAM. The computation scripts are developed within MATLAB, using the Yalmip and Gurobi toolboxes.
	\subsection{Test System and Data}
	The dichotomy and FMCS method (D\&FMCS) is tested using the RBTS and RTS-79 system. RBTS is a small-scale power system with 11 generations, 9 transmission lines and 6 buses. The total installed capacity is 240 MW, with an annual peak load of 185 MW. RTS-79 is a composite power system with 33 generators, 38 transmission lines and 24 buses. The total installed capacity is 3405 MW, with an annual peak load of 2850 MW.
	
	\subsection{Result and Analysis}
	The stopping criterion for the dichotomy method is set as the average of the probabilities of the last ten evaluated failed lattices. In addition, we denote \( 1\times 10^{-n} \) by \(\text{D}_n\).
	
	\emph{ 1) RBTS System}
	\begin{figure}[h]
		\centering
		\includegraphics[width=3in]{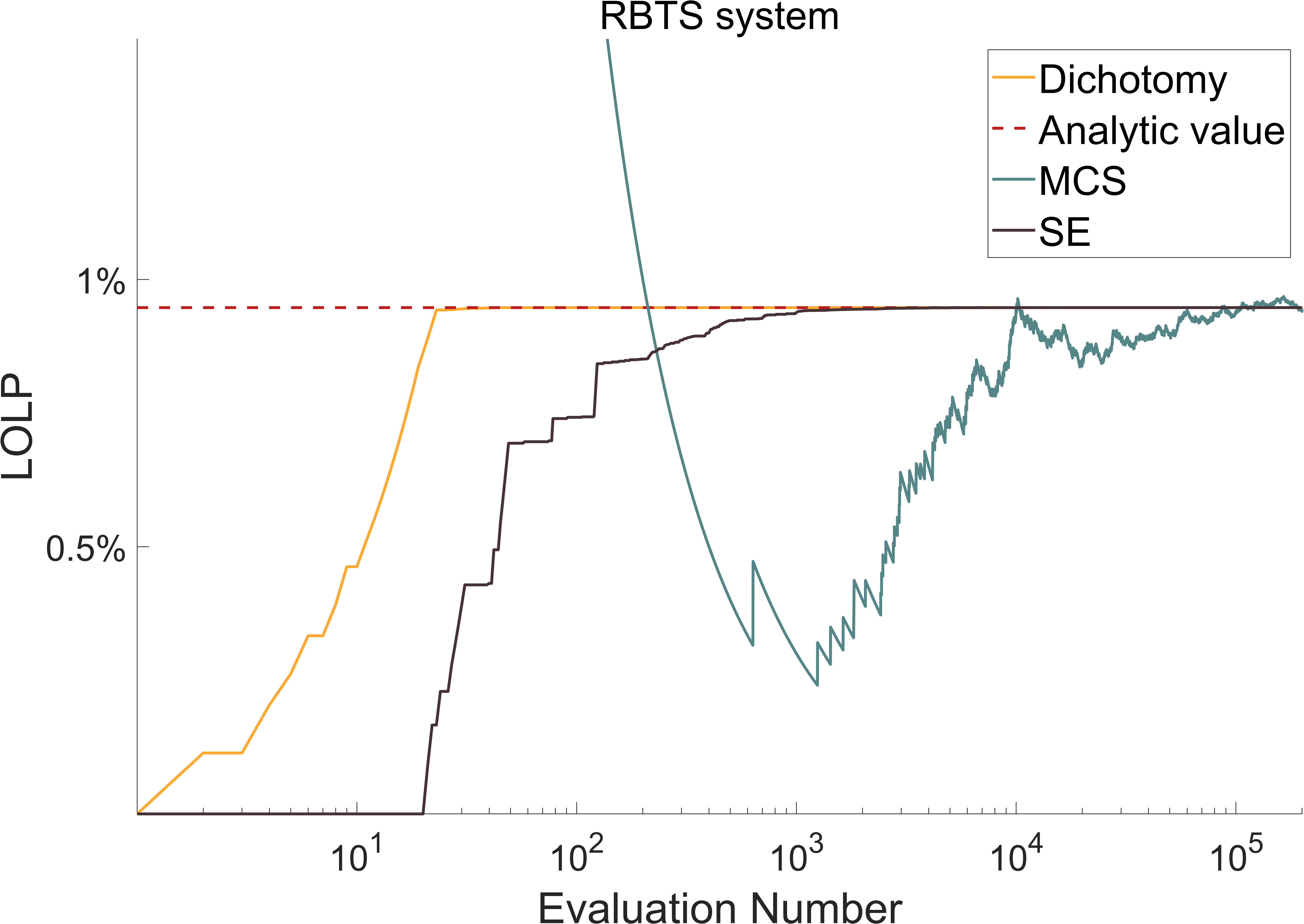}
		\caption{Comparison of LOLP Calculated by Different Methods(RBTS)}
		\label{RBTSLOLP}
	\end{figure}
	
	\begin{figure}[ht]
		\centering
		\includegraphics[width=3in]{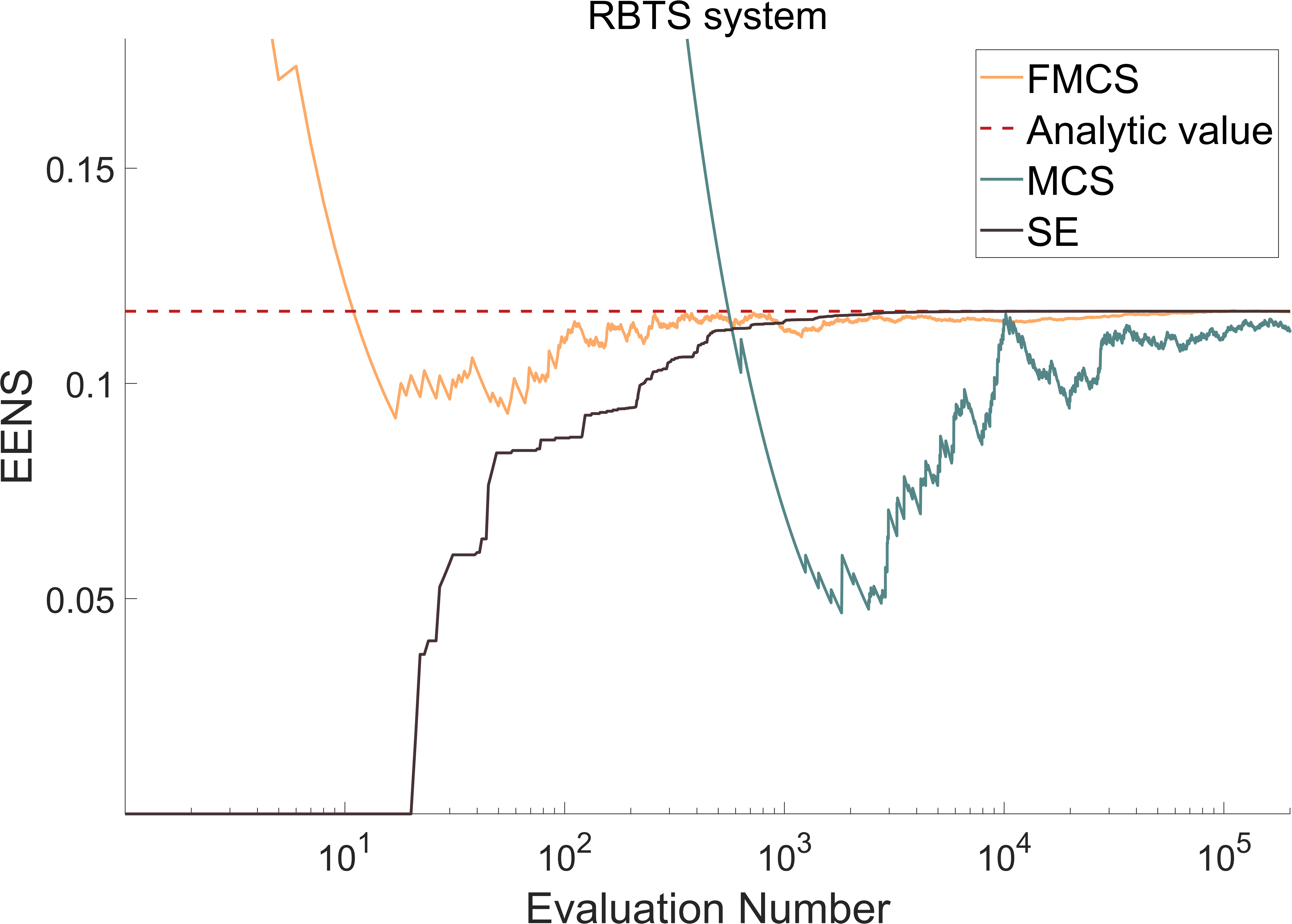}
		\caption{Comparison of EENS Calculated by Different Methods(RBTS)}
		\label{RBTSEENS}
	\end{figure}
	\begin{table}[h]
		\centering
		\caption{Failed State by SE(RBTS)}
		\begin{tabular}{|c|c|c|}
			\hline
			1-level & 2-level & 3-level \\
			\hline
			\multirow{10}*{\{20\}} & \{1,2\},\{1,4\},\{1,7\},\{1,8\},\{1,9\}, & \{1,2,3\},\{1,2,4\}, \\ 
			& \{1,10\},\{1,11\},\{1,20\}, & \{1,2,5\},\{1,2,6\}, \\ 
			& \{2,4\},\{2,7\},\{2,8\},\{2,9\}, & \{1,2,7\},\{1,2,8\}, \\
			& \{2,10\},\{2,11\},\{2,20\},\{3,20\}, & \{1,2,11\},\{1,2,12\}, \\
			& \{4,7\},\{4,20\},\{5,20\},\{6,20\}, & \{1,2,13\},\{1,2,15\}, \\ 
			& \{7,8\},\{7,9\},\{7,10\},\{7,11\}, & \{1,2,16\},\{1,2,17\}, \\
			& \{7,20\},\{8,20\},\{9,20\},\{10,20\}, & \{1,2,19\},\{1,2,20\},\\
			& \{11,20\},\{12,20\},\{13,20\},\{14,20\}, & \{1,3,4\}, \{1,3,7\} \\
			& \{15,20\},\{16,19\},\{16,20\},\{17,20\}, & \{1,3,8\}, \{1,3,9\}\\
			& \{18,20\},\{19,20\} & \{1,3,10\}, \(\cdots\)\\
			\hline
		\end{tabular}
	\end{table}
	\begin{table}[ht]
		\centering
		\caption{Failed Lattices by Dichotomy(RBTS)}
		\begin{tabular}{ccccc}
			\toprule
			\( \mathcal{L}^F \)  & \( \hat{0} \) & \( C\left(\hat{0}\right) \) /MW & Num. of states & \( P\left(\mathcal{L}^F\right) \) (Eq.13) \\
			\midrule
			1 & \{20\} & 20 & \(2^{19}\) & 0.11402\% \\
			2 & \{1,2\} & 25 & \(2^{17}\) & 0.08990\% \\
			3 & \{1,7\} & 25 & \(2^{16}\) & 0.05813\% \\
			4 & \{1,4\} & 5 & \(2^{15}\) & 0.07121\% \\
			5 & \{2,7\} & 25 & \(2^{16}\) & 0.05813\% \\
			6 & \{2,4\} & 5 & \(2^{15}\) & 0.07121\% \\
			7 & \{7,4\} & 5 & \(2^{15}\) & 0.04699\% \\
			8 & \{2,8\} & 5 & \(2^{14}\) & 0.04166\% \\
			9 & \{1,8\} & 5 & \(2^{14}\) & 0.04166\% \\
			10 & \{1,9\} & 5 & \(2^{13}\) & 0.04104\% \\
			\vdots & \vdots & \vdots & \vdots & \vdots \\
			15 & \{2,11\} & 5 & \(2^{11}\) & 0.03981\% \\
			16 & \{7,8\} & 5 & \(2^{14}\) & 0.02749\% \\
			17 & \{7,9\} & 5 & \(2^{13}\) & 0.02708\% \\
			18 & \{7,10\} & 5 & \(2^{12}\) & 0.02667\% \\
			19 & \{7,11\} & 5 & \(2^{11}\) & 0.02627\% \\
			20 & \{4,8,9\} & 5 & \(2^{13}\) & 0.00051\% \\
			\vdots & \vdots & \vdots & \vdots & \vdots \\
			30 & \{16,19\} & 40 & \(2^{11}\) & 0.00011\% \\
			\bottomrule
		\end{tabular}
	\end{table}
	As shown in Table \uppercase\expandafter{\romannumeral1}, SE enumerates all \(2^{20}=1,048,576\) states to calculate the analytical values of the indices, which serves as the benchmark for other methods. When taking \(n=7\), the dichotomy method demonstrates high precision and efficiency in calculating LOLP, achieving an error of 0.0007\% after performing OPF on 190 states. Within the failed space identified by the dichotomy method, FMCS is used to calculate EENS. With \( \beta_{\text{EENS}} = 0.01 \), there are 6,874 states are sampled, resulting in an error of 0.1382\%. The MCS method also exhibits high accuracy, with an error of 0.0008\% for LOLP and 0.3880\% for EENS under the same \(\beta_{\text{EENS}}\). However, the FMCS method is approximately 250 times faster than MCS. This is because the direct MCS samples a large number of normal states for the high stability of RBTS, while FMCS always samples failed stats within the failed space.
	
	In particular, the convergence rate of the FMCS method is almost the same when taking \( n \) as 7 and 9 respectively. The higher the \( n \) sets, the more precise of the failed space is.
	
	Fig.3 and Fig.4 show the changes in LOLP and EENS with the number of OPF evaluations using the SE, MCS and D\&FMCS method. As illustrated in Fig.3, when calculating LOLP, both the dichotomy and SE converge towards the analytic value in a monotonically increasing fashion. Regarding the rate at which the curves reach stability, the dichotomy method is approximately 100 times faster than the SE and about 10,000 times faster than the MCS. As illustrated in Fig.4, when calculating EENS, the SE and FMCS approach stability almost concurrently. The FMCS is approximately 100 times faster than the MCS. This phenomenon is attributed to the small scale and high stability of the RBTS system, which allows the enumeration method to be highly efficient, while the MCS struggles to converge rapidly.

	Table \uppercase\expandafter{\romannumeral2} presents the failed states of 1-level, 2-level and a part of 3-level of RBTS, as determined by SE. Table \uppercase\expandafter{\romannumeral3} illustrates the partial failed lattices, as determined by the dichotomy method.
	
	A comparative analysis of \uppercase\expandafter{\romannumeral2} and \uppercase\expandafter{\romannumeral3}, \(\{1,20\}\), \(\{2,20\}\), \(\cdots\), \(\{18,20\}\) and \(\{19,20\}\) , are all encompassed within the failed lattice \([\{20\},\{1,2,\cdots,20\}]\). Similarly, the failed states of the 3-level that include component 20 are also contained within this lattice.  As depicted in Table \uppercase\expandafter{\romannumeral3}, the minimum element of the 20th failed lattice is of the 3-level, while the 30th failed lattice is of the 2-level. Consequently, the level of minimum elements within the partitioned failed lattice does not strictly increase, instead, they are predominantly influenced by the importance of the components and lattices.
	
	\emph{ 2)  RTS-79 System}
	
	\begin{table*}[ht]
		\centering
		\caption{RELIABILITY ASSESSMENT RESULTS OF SE, MCS, and D-FMCS (RTS-79)}
		\begin{tabular}{ccccccccc}
			\toprule
			\multicolumn{2}{c}{Method} & \(\beta_{\text{EENS}}\) & LOLP & LOLP error & EENS & EENS error & Evaluation Number & CPU time(s) \\
			\midrule
			\multicolumn{2}{c}{MCS} & --& 8.47569\% & 0.0000\% & 14.7941 & 0.0000\% & 1,000,000 & 62,143.2 \\
			\midrule
			\multicolumn{2}{c}{SE} & -- & 7.62093\% & 10.085\% & 12.1306 & 18.004\% & 974,121 & 61,101.7 \\
			\midrule
			\multicolumn{2}{c}{\multirow{3}*{MCS}} 
			& 0.1 & 9.59072\% & 13.156\% & 17.1002 & 15.588\% & 1,637 & 106.0 \\
			& & 0.05 & 8.81370\% & 3.9879\% & 16.0481 & 8.4763\% & 7,477 & 468.4 \\
			& & 0.01 & 8.40650\% & 0.8164\% & 14.6012 & 1.3043\% & 195,337 & 12,210.0 \\
			\midrule
			\multirow{4}{*}{D-FMCS}& Dichotomy (\(\text{D}_7\))
			& -- & 8.45786\% & 0.2104\% & -- & -- & 6,319 & 398.7 \\ 
			\cmidrule{2-9}
			&\multirow{3}*{FMCS}
			& 0.1  & -- & -- & 14.1547 & 4.3220\% & 73 & 4.7 \\
			& & 0.05 & -- & -- & 15.1081 & 2.1222\% & 277 & 17.4 \\
			& & 0.01 & -- & -- & 14.7472 & 0.3173\% & 7,497 & 467.9 \\
			\midrule
			\multirow{4}{*}{D-FMCS} & Dichotomy (\(\text{D}_9\))
			& -- & 8.46323\% & 0.1470\% & -- & -- & 29,066 & 1,852.7 \\ 
			\cmidrule{2-9}
			&\multirow{3}*{FMCS}
			& 0.1 & -- & -- & 14.7143 & 0.5396\% & 94 & 5.9 \\
			& & 0.05 & -- & -- & 15.5285 & 4.9639\% & 305 & 19.1 \\
			& & 0.01 & -- & -- & 14.7746 & 0.1322\% & 7,177 & 448.9 \\
			\bottomrule
		\end{tabular}
	\end{table*}
	\begin{figure}[h]
		\centering
		\includegraphics[width=3in]{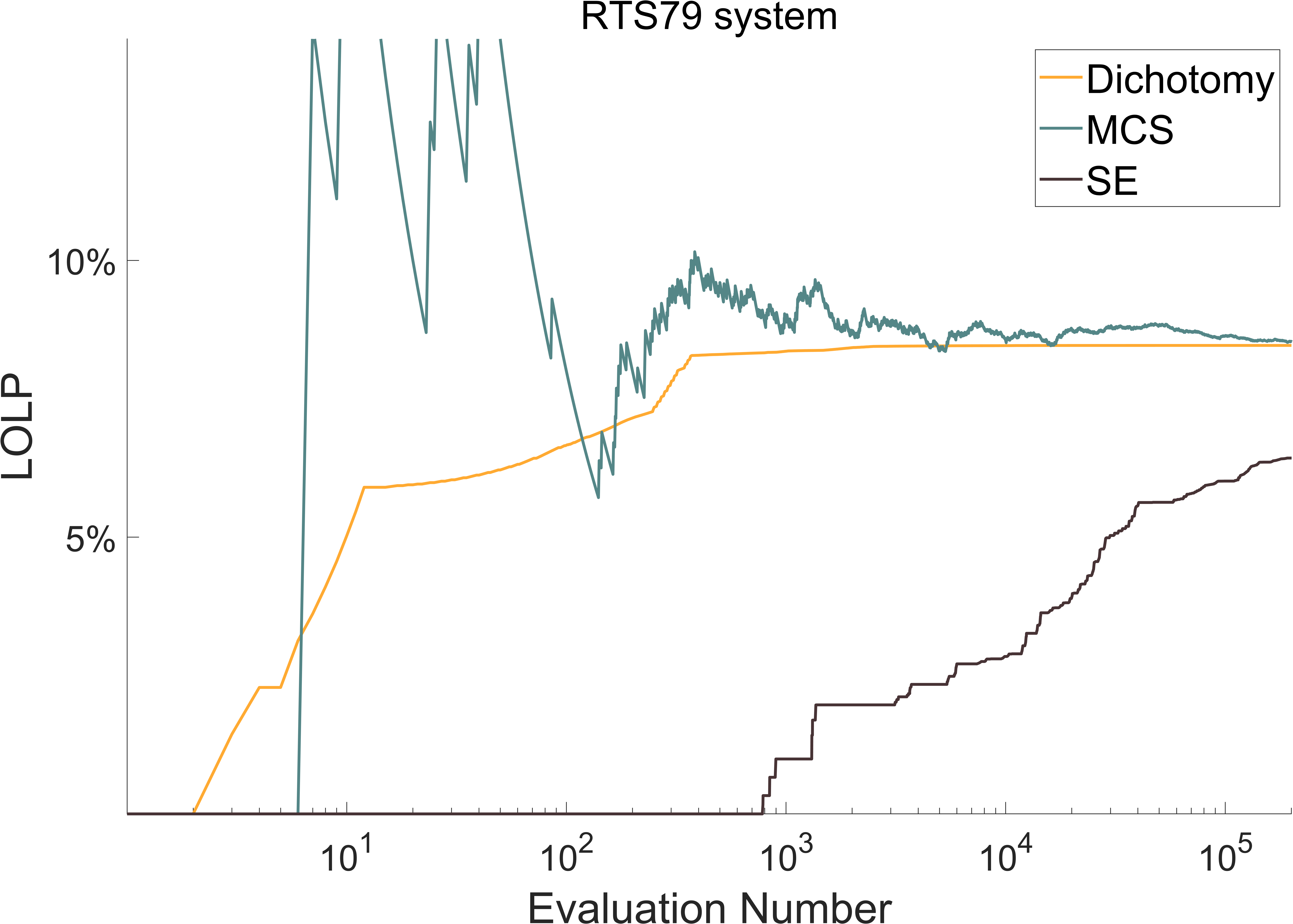}
		\caption{Comparison of LOLP Calculated by Different Methods(RTS-79)}
		\label{RTSLOLP}
	\end{figure}
	\begin{figure}[h]
		\centering
		\includegraphics[width=3in]{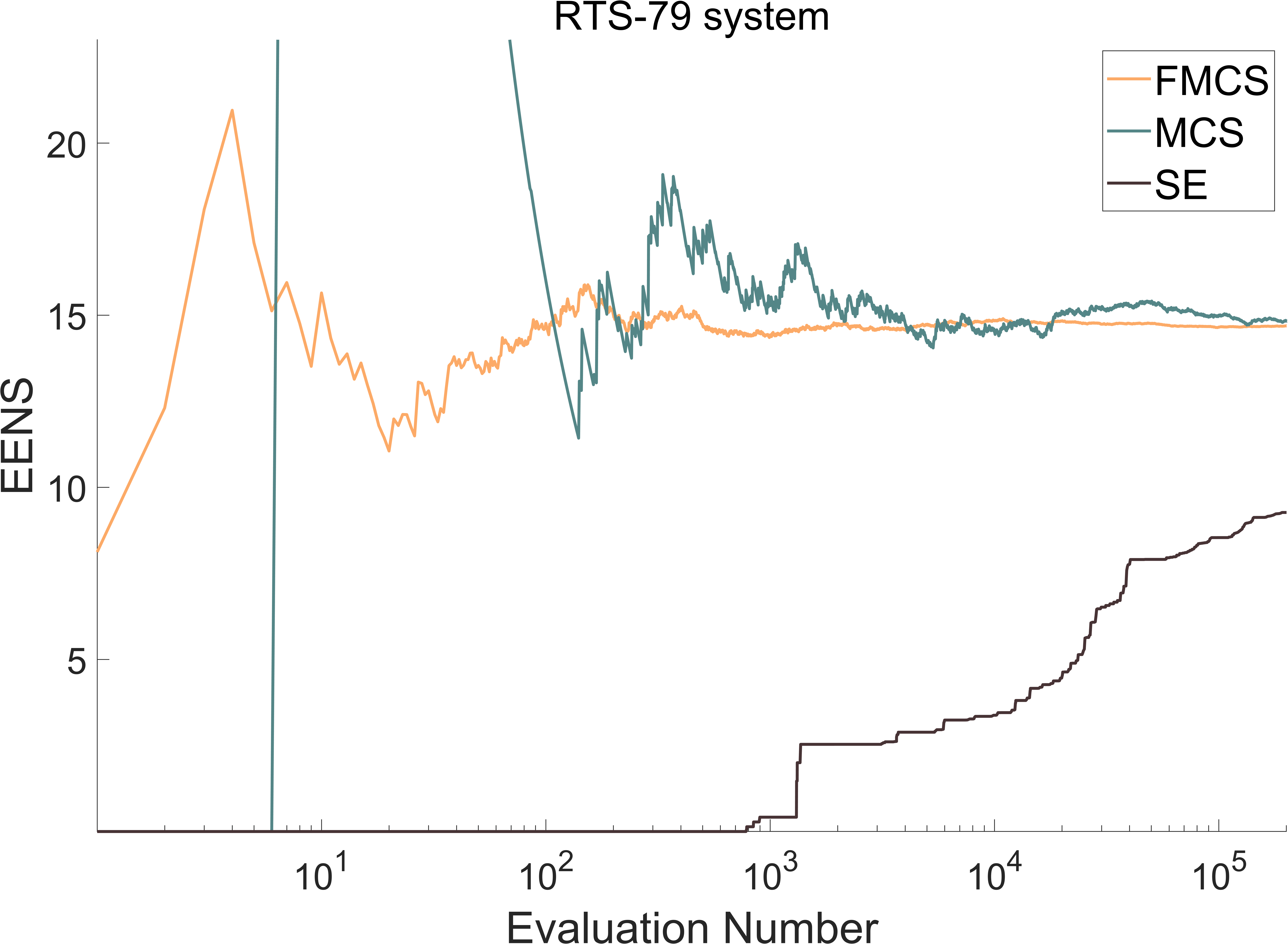}
		\caption{Comparison of EENS Calculated by Different Methods(RTS-79)}
		\label{RTSEENS}
	\end{figure}
	Table \uppercase\expandafter{\romannumeral4} illustrates the reliability indices of the RTS-79 system as determined by the SE, MCS, and D\&FMCS method. MCS samples 1,000,000 states to calculate the index as the benchmark value. SE comprehensively enumerates all contingency states up to the 4th level, totaling \( \binom{70}{0} + \binom{70}{1} + \binom{70}{2} + \binom{70}{3} + \binom{70}{4} = 974,121 \) states. With \( \beta_{\text{EENS}} = 0.01 \), MCS computes a LOLP error of 0.8164\% and an EENS error of 1.3043\%. Under the same coefficient of variation and \(n=7\), D\&FMCS yields a LOLP error of 0.2104\% and an EENS error of 0.3173\%. When compared under the same convergence condition, \(\beta = 0.01\), F\&FMCS is approximately \( \frac{12210.0}{398.7 + 467.9} \approx 14\) times faster than MCS, and the calculation error is smaller. When \(n\) increases to 9, the accuracy of LOLP improves, with the error reducing to 0.1470\%. Conceptually, the dichotomy method provides a lower bound of LOLP, whereas MCS offers a stochastic estimate. If the LOLP determined by MCS were to fall below that of the dichotomy method, it would cast doubt on the validity of the MCS results.
	
	Fig.5 and Fig.6 depict the convergence of reliability index for the RTS-79 system calculated using various methods. The dichotomy method and FMCS exhibit faster convergence compared to MCS. When comparing Fig.3 with Fig.5, it is evident that the efficiency of SE is significantly decreased, and similarly, the efficiency of the dichotomy method also decreases. This suggests that the efficiency of the dichotomy method is influenced by the size of the state space. Given that the state space scales from \( 2^{20} \) to \( 2^{70} \), the reduction in the efficiency of the dichotomy method is considered acceptable.
	\begin{table}[h]
		\centering
		\caption{Failed State by SE(RTS-79)}
		\begin{tabular}{|c|c|c|}
			\hline
			1-level & 2-level & 3-level \\
			\hline
			\multirow{5}*{None} & \{12,22\},\{12,23\},\{13,22\}, & \{1,12,22\},\{1,12,23\}, \\ 
			& \{13,23\},\{14,22\},\{14,23\}, & \{1,12,32\},\{1,13,22\}, \\ 
			& \{22,23\},\{22,32\},\{22,43\}, & \{1,13,23\},\{1,13,32\}, \\
			& \{23,32\},\{23,43\},\{35,41\}, & \{1,14,22\},\{1,14,23\}, \\
			& \{36,40\},\{37,42\},\{51,55\} & \{1,14,32\},\(\cdots\) \\
			\hline
		\end{tabular}
	\end{table}
	
	Table \uppercase\expandafter{\romannumeral5} shows the failed states of 1-level, 2-level and a part of 3-level of RTS-79 system, as determined by SE. 
	
	Table \uppercase\expandafter{\romannumeral6} presents the partial failed lattices determined by the dichotomy method. The components with installed capacities of 400 MW (components 22 and 23), 350 MW (component 32), and 197 MW each (components 12, 13, and 14) are particularly critical, as their failures are more likely to cause a system-wide power shortage. Focusing on the 299th failed lattice, the failure of component 43 results in the disconnection of node 7 from the power grid. This indicates that the partitioning of failed lattices corresponds to the importance of components and lattices, offering substantial support for the reliability analysis of the power system.
	
	\begin{table}[H]
		\centering
		\caption{Failed Sublattices by Dichtomy(RTS-79)}
		\begin{tabular}{ccccc}
			\toprule
			\( \mathcal{L}^F \)  & \( \hat{0} \) & \( C\left(\hat{0}\right) \)(MW) & Num. of states  & \( P\left(\mathcal{L}^F\right) \) (Eq.13) \\
			\midrule
			1 & \{22,23\} & 245 & \( 2^{68} \) & 1.44000\% \\
			2 & \{22,32\} & 195 & \( 2^{67} \) & 0.84480\%\\
			3 & \{23,32\} & 195 & \( 2^{67} \) & 0.84480\%\\
			4 & \{23,12\} & 42 & \( 2^{66} \) & 0.48576\% \\
			5 & \{22,12\} & 42 & \( 2^{66} \) & 0.48576\% \\
			6 & \{22,13\} & 42 & \( 2^{65} \) & 0.46147\% \\
			7 & \{23,13\} & 42 & \( 2^{65} \) & 0.46147\% \\
			8 & \{23,14\} & 42 & \( 2^{64} \) & 0.43840\% \\
			9 & \{22,14\} & 42 & \( 2^{64} \) & 0.43840\% \\
			10 & \{32,12,13\} & 189 & \( 2^{64} \) & 0.01548\% \\
			11 & \{32,12,14\} & 189 & \( 2^{63} \) & 0.01471\% \\
			\vdots & \vdots & \vdots & \vdots & \vdots \\
			298 & \{22,31,19\} & 12 & \( 2^{38} \) & 0.00274\% \\
			299 & \{23,43\} & 20 & \( 2^{56} \) & 0.00214\% \\
			300 & \{22,43\} & 20 & \( 2^{56} \) & 0.00214\% \\
			301 & \{32,12,24\} & 42 & \( 2^{51} \) & 0.00194\% \\
			\bottomrule
		\end{tabular}
	\end{table}
	\begin{figure}[!ht]
		\centering
		\includegraphics[width=3.1in]{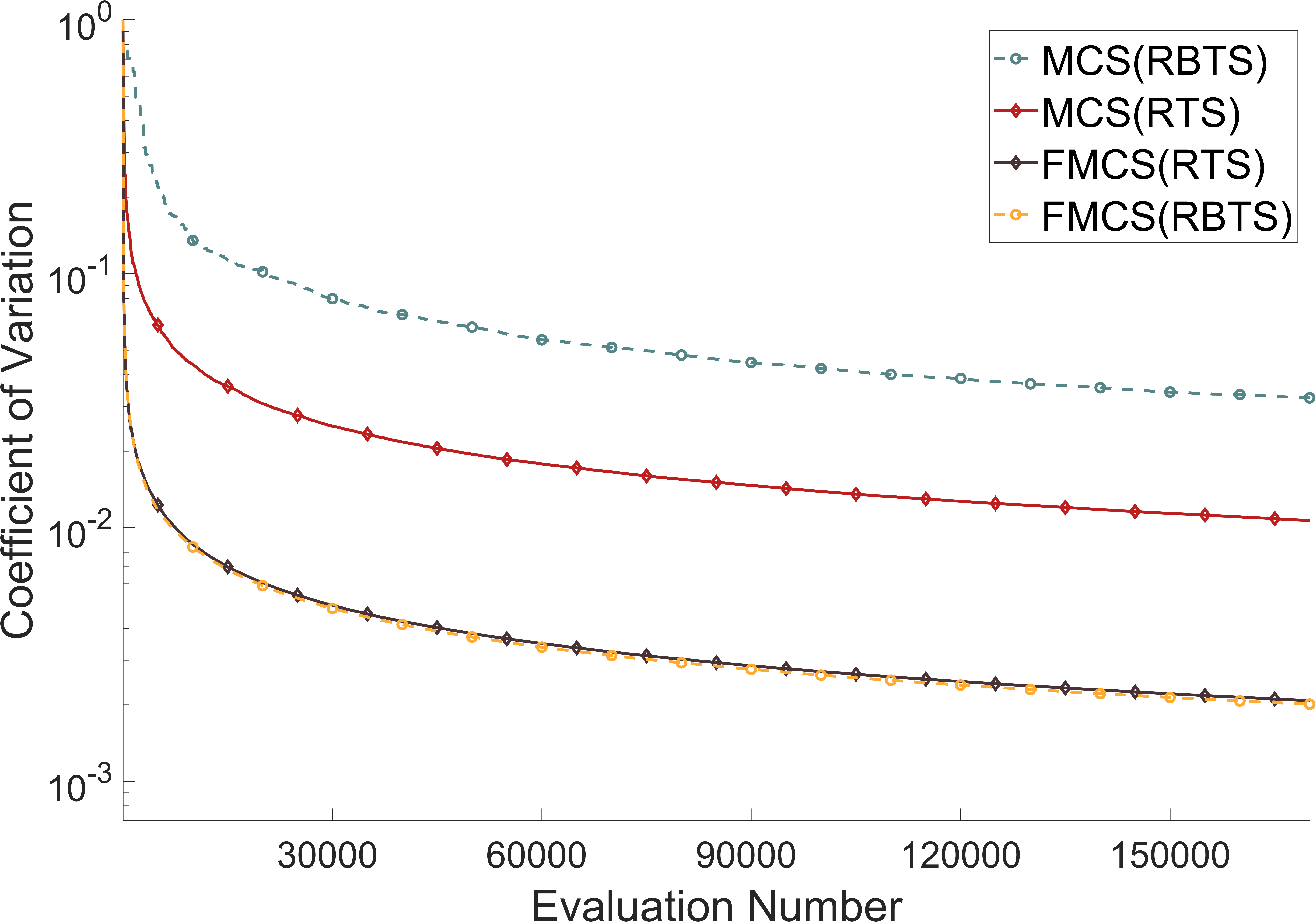}
		\caption{Evolution of the coefficient of variation for EENS.}
		\label{VF}
	\end{figure}
	\subsection{Analysis of \(\beta_{\text{EENS}}\) and Computing Time}
	Fig.7 illustrates the evolution of the coefficient of variation for EENS evolves as the number of OPF evaluation increases. MCS is sensitive to system stability, resulting in a marked disparity in the rate of change of the coefficient of variation for EENS between the RBTS and RTS-79 systems. In contrast, FMCS is not influenced by the stability of the power system and exhibits a nearly uniform decrease in the EENS COV across both systems, which is also more rapid than that observed with MCS. Consequently, if the failure space can be accurately determined, the convergence rate of the FMCS will remain unaffected by the size and stability of the system.
	
	In the experimental result, the computational time for the dichotomy method was predominantly attributed to the OPF calculations and the prioritization of component selection, which accounted for 98.4\% and 0.6\% of the total computation time, respectively.
	
	\section{Conclusion}
	In this paper, we present a novel space partitioning technique called the dichotomy method, which calculates reliability indices lattice-by-lattice. The dichotomy method partitions the state space into failed and normal space based on the Cartesian product and lattices categories. In addition, MCS can be applied directly in the failed space because of the unique structure of a lattice. The results show that the dichotomy method effectively partitions the state space by evaluating a limited number of states. The FMCS method estimates EENS through sampling in the failed space, with its convergence rate not affected by the size or stability of the system. Moreover, the minimum element of the failed lattice plays a crucial role in identifying the power system's vulnerabilities. The experimental results from both the RBTS and RTS systems indicate that the dichotomy method and FMCS compute reliability indices faster and more accurately, confirming the viability of our proposed methodologies.
	
	In the dichotomy method, space partitioning alternates with state evaluation. The computation time primarily arises from the OPF calculation. Our future work will explore the relationship between space partitioning and state evaluation, which is key to reducing the computational time of the dichotomy method.
	\bibliographystyle{IEEEtran}
	\bibliography{ref}
\end{document}